# Nonlinear lattice dynamics as a basis for enhanced superconductivity in YBa$_2$Cu$_3$O$_{6.5}$

R. Mankowsky[1*], A. Subedi[2*], M. Först[1], S.O. Mariager[3], M. Chollet[4], H. T. Lemke[4], J. S. Robinson[4], J. M. Glownia[4], M. P. Minitti[4], A. Frano[5], M. Fechner[6], N. A. Spaldin[6], T. Loew[5], B. Keimer[5], A. Georges[2,7,8] and A. Cavalleri[1,9,10]

[*] These authors contributed equally to this work.
[1] Max Planck Institute for the Structure and Dynamics of Matter, 22761 Hamburg, Germany
[2] Centre de Physique Théorique, École Polytechnique, CNRS, 91128 Palaiseau Cedex, France
[3] Swiss Light Source, Paul Scherrer Institut, 5232 Villigen, Switzerland
[4] Linac Coherent Light Source, SLAC National Accelerator Laboratory, Menlo Park 94025, California, USA
[5] Max Planck Institute for Solid State Research, 70569 Stuttgart, Germany
[6] Eidgenössische Technische Hochschule Zürich, Materials Theory, 8093 Zürich, Switzerland
[7] Collège de France, 11 place Marcelin Berthelot, 75005 Paris, France
[8] Département de Physique de la Matière Condensée MaNEP, Université de Genève, 1211 Genève, Switzerland
[9] Department of Physics, Oxford University, Clarendon Laboratory, Oxford OX1 3PU, UK
[10] Center for Free-Electron Laser Science CFEL and University of Hamburg, 22761 Hamburg, Germany

**THz-frequency optical pulses can resonantly drive selected vibrational modes in solids and deform their crystal structure[1,2,3]. In complex oxides, this method has been used to melt electronic orders[4,5,6], drive insulator to metal transitions[7,8] or induce superconductivity[9]. Strikingly, coherent interlayer transport strongly reminiscent of superconductivity can be transiently induced up to room temperature in YBa$_2$Cu$_3$O$_{6+x}$[10,11]. By combining femtosecond X-ray diffraction and *ab initio* density functional theory calculations, we determine here the crystal structure of this exotic non-equilibrium state. We find that nonlinear lattice excitation in normal-state YBa$_2$Cu$_3$O$_{6+x}$ at 100 K causes a staggered dilation/contraction of the Cu-O$_2$ intra/inter-bilayer distances, accompanied by anisotropic changes in the in-plane O-Cu-O bond buckling. Density functional theory calculations indicate that these motions cause**

**dramatic changes in the electronic structure. Amongst these, the enhancement in the $d_{x2-y2}$ character of the in-plane electronic structure is likely to favor superconductivity.**

The response of a crystal lattice to strong, resonant excitation of an infrared-active phonon mode can be described by separating the crystal Hamiltonian into its linear and nonlinear terms $H=H_{lin}+H_{NL}$. The linear term $H_{lin} = \frac{1}{2}\omega_{IR}^2 Q_{IR}^2$ describes harmonic oscillations about the equilibrium atomic positions, with $\omega_{IR}$ denoting the frequency and $Q_{IR}$ the normal coordinate of the infrared-active mode. In the limit of lowest order (cubic) coupling to other modes with generic coordinate $Q_R$, the nonlinear term can be written as $H_{NL} = \frac{1}{2}\omega_R^2 Q_R^2 - a_{12}Q_{IR}Q_R^2 - a_{21}Q_{IR}^2 Q_R$. In this expression, $a_{12}$ and $a_{21}$ are anharmonic coupling constants. (See online Methods section and Extended Data Figures 1 and 2 for details on next order coupling). For a centrosymmetric crystal like YBa$_2$Cu$_3$O$_{6.5}$, $Q_{IR}Q_R^2$ is zero as $Q_{IR}$ is odd in symmetry while $Q_R^2$ is even. Furthermore, as $Q_{IR}^2$ is of even symmetry $Q_{IR}^2 Q_R$ is nonzero only if $Q_R$ is even and hence Raman active.

Thus, the total Hamiltonian reduces to $H = \frac{1}{2}\omega_{IR}^2 Q_{IR}^2 + \frac{1}{2}\omega_R^2 Q_R^2 - a_{21}Q_{IR}^2 Q_R$, which predicts a shift in the potential energy minimum along $Q_R$ for any finite distortion $Q_{IR}^*$ (see Figure 1a). Correspondingly, for a periodically driven $Q_{IR}$ mode, the dynamics are described by the coupled equations of motion $\ddot{Q}_{IR} + 2\gamma_{IR}\dot{Q}_{IR} + \omega_{IR}^2 Q_{IR} = f(t) + 2a_{21}Q_{IR}Q_R$ and $\ddot{Q}_R + 2\gamma_R\dot{Q}_R + \omega_R^2 Q_R = a_{21}Q_{IR}^2$.

Figure 1b pictorially represents these dynamics. Upon resonant mid-infrared excitation of $Q_{IR}$, a *unidirectional* force is exerted along the normal coordinate $Q_R$, which is *displaced* by a magnitude proportional to $Q_{IR}^2$. This effect remains sizeable only as long as $Q_{IR}$ oscillates coherently, typically for several picoseconds.

We next discuss the specific case of YBa$_2$Cu$_3$O$_{6.5}$, which crystallizes in a centrosymmetric orthorhombic unit cell with D$_{2h}$ symmetry, comprising bilayers of conducting CuO$_2$ planes separated by an insulating layer containing Yttrium atoms and Cu-O chains that control the doping of the planes (Fig. 2a). The YBa$_2$Cu$_3$O$_{6.5}$ sample contained both oxygen rich and oxygen deficient chains, and exhibited short-range Ortho-II ordering of the vacancies (Fig. 2b). Note also that the in-plane O-Cu-O bonds are buckled (Fig. 2c).

In our experiments, mid-infrared pump pulses of ~300 femtoseconds (fs) duration were focused to a maximum fluence of ~4 mJ/cm$^2$ and a peak electric field of ~3 MV/cm. These pulses were polarized along the *c* axis of YBa$_2$Cu$_3$O$_{6.5}$ and tuned to resonance with the same 670 cm$^{-1}$ frequency (~15 μm, 83 meV) B$_{1u}$ infrared-active mode[12] (see Figure 2a) that was previously shown by means of time-resolved THz spectroscopy to enhance interlayer superconducting coupling[10,11].

In analyzing the nonlinear lattice dynamics caused by this excitation, we note that the nonlinear term $Q_{B1u}^2 Q_R$ is non zero only if $Q_R$ is of A$_g$ symmetry, because the square of the irreducible representation of B$_{1u}$ is A$_g$. Thus, only A$_g$ modes can couple to the optically driven B$_{1u}$ motion of Figure 2a. YBa$_2$Cu$_3$O$_{6.5}$ has 72 optical phonon modes, of which 33 are Raman active. These can be further divided into 22 B$_g$ modes, which break in-plane symmetry, and 11 A$_g$ modes, which preserve the symmetry of the unit cell (see Extended Data Figure 3). The geometries of these 11 A$_g$ modes and their coupling strength to the driven B$_{1u}$ mode were computed using first principles density functional theory calculations within the local density approximation. At 3 MV/cm field strength of the mid-infrared pump pulse, we expect a peak amplitude for the B$_{1u}$ motion corresponding to 2.2pm increase in apical oxygen-Cu distance, which was used as a basis to calculate energy potentials of the A$_g$ modes for a frozen distortion of this magnitude (see Figure 3a). Only four phonon modes A$_g$(15, 21, 29,

74) were found to strongly couple to the driven $B_{1u}$ mode, all involving a concerted distortion of the apical oxygen atoms towards the $CuO_2$ plane and an increase in Cu-O buckling (Figure 3b). The calculations also predict weak coupling to three further modes $A_g$(52, 53, 61), consisting of breathing motion of the oxygen atoms in the plane (Fig. 3c). The remaining four modes $A_g$(14, 39, 53, 65) do not couple to the $B_{1u}$ mode (see Extended Data Table 1 for details).

To experimentally determine the absolute amplitude of these distortions in the conditions relevant for enhanced superconductivity[10,11], we measured time resolved x-ray diffraction using 50 fs, 6.7 keV pulses from the LCLS free electron laser, which was synchronized to the optical laser that generated the mid-infrared pump pulses. Changes in diffraction intensity were recorded for four Bragg peaks at a base temperature of 100 K, above the equilibrium transition temperature $T_c$ = 52 K. These peaks were observed to either increase or decrease promptly after excitation (see Figure 4) and to relax within the same timescale as the changes in the THz optical properties[10,11]. For each Bragg reflection we calculated changes in diffraction as function of $B_{1u}$ amplitude considering a displacement of only the four dominant Raman modes $A_g$(15, 21, 29, 74) or all 11 modes of Figure 3, taking into account the relative coupling strengths. We simultaneously fitted the four experimental diffraction curves using only two free parameters: the amplitude of the directly driven $B_{1u}$ motion and the relative contributions of two exponential relaxation components ($\tau_1$ = 1 ps, $\tau_2$ = 7 ps) extracted from the THz measurements[10,11]. Very similar results were found when considering only the four dominant modes or all modes (see green and red fitting curves in Figure 4).

The transient lattice structure determined from these fits involves the following elements. Firstly, we observe a *decrease* in the distance between the apical oxygen and the copper atoms of the superconducting planes (Figure 5). This motion is far smaller and opposite in sign than the difference in the static apical oxygen positions between La- and Hg-based cuprates, for which $T_c$ increases at equilibrium[13]. Therefore, the transient enhancement of superconducting transport cannot be explained by this analogy. More suggestively, the copper atoms are driven away from one another within the bilayers and toward one another between different bilayers. This staggered motion is of approximately 0.63% (see Figure 5a) and qualitatively follows the decrease in intra-bilayer tunneling and the enhancement of inter-bilayer tunneling[10,11]. Finally, an anisotropic 0.32° increase in the in-plane O-Cu-O buckling (different along a and b axes) is observed (see Extended Data Table 2).

Although the Josephson coupling in layered cuprates involves many microscopic parameters that are not kept into account here [14,15,16], DFT calculations in the distorted crystal structure were used to assess the salient effects on the electronic properties (see Extended Data Figures 4 to 6). Our calculations predict an energy lowering of the oxygen deficient chain bands by few tens of meV. Because at equilibrium these bands are very close to the Fermi level, this small shift strongly reduces the hybridization of the chains with the plane Cu-orbital, leading to a DFT Fermi surface with a stronger Cu $d_{x^2-y^2}$ character and higher hole doping. This effect is likely to favor superconductivity. We also speculate that as the DFT Fermi surface changes shape and size, it is quite possible that that charge density wave order may also be destabilized[17,18,19], which would also aid superconductivity. The present calculations will serve as a starting point for a full many-body treatment, to be complemented by more exhaustive experimental characterizations of the transient electronic structure.

More generally, we see nonlinear phononics as a new tool for dynamical materials discovery, with optical lattice control providing a perturbation -- analogous to strain, fields or pressure

-- that can induce exotic collective electronic behaviors. Knowledge of the non-equilibrium atomic structure by ultrafast x-ray crystallography, which we provide for the first time in this work, is the essential next step towards engineering these induced behaviors at equilibrium.

**References Main Text**

## Acknowledgements

The research leading to these results has received funding from the European Research Council under the European Union's Seventh Framework Programme (FP7/2007-2013) / ERC Grant Agreement n° 319286 (QMAC). Funding from the priority program SFB925 of the German Science Foundation (DFG) is gratefully acknowledged.

Portions of this research were carried out at the Linac Coherent Light Source (LCLS) at the SLAC National Accelerator Laboratory. LCLS is an Office of Science User Facility operated for the U.S. Department of Energy Office of Science by Stanford University. This work was supported by the Swiss National Supercomputing Centre (CSCS) un- der project ID s404.

This work was supported by the Swiss National Science Foundation through its National Center of Competences in Research MUST.

## Author Contributions

A.C. conceived this project. R.M. and M.F. lead the diffraction experiment, supported by S.O.M., M.C., H.T.L., J.S.R., J.M.G., M.P.M., and A.F.. R.M. and A.S. analyzed the data. A.S. performed the DFT calculations, with support from A.G., M.F. and N.A.S.. The sample was grown by T.L and B.K.. R.M. and A.C. wrote the manuscript, with feedback from all co-authors.

## Author Information

Correspondence and requests for materials should be addressed to Andrea Cavalleri (andrea.cavalleri@mpsd.mpg.de) or Roman Mankowsky (roman.mankowsky@mpsd.mpg.de)

# Figures

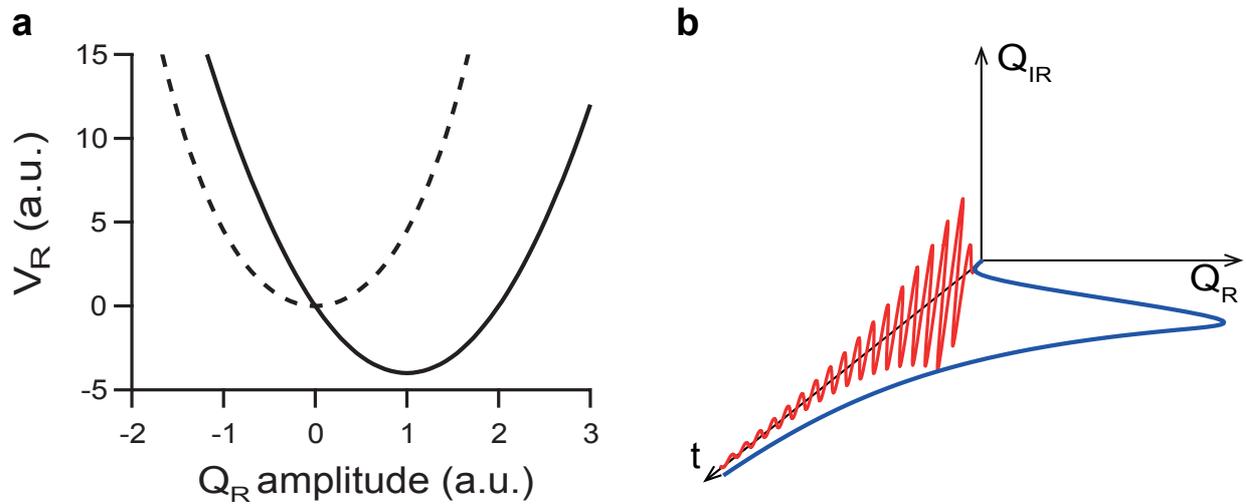

**Figure 1 | Coherent nonlinear lattice dynamics in the limit of cubic coupling. a)** A static distortion $Q_{IR}^*$ shifts the equilibrium potential (dashed line) of all modes $Q_R$ that are coupled through $Q_{IR}^2 Q_R$ coupling, displacing the equilibrium position towards a new minimum (solid line). **b)** The dynamical response of the two modes involves an oscillatory motion of the infrared mode (red line) and a *directional* displacement of $Q_R$ (blue line). The displacement is proportional to $Q_{IR}^2$ and survives as long as $Q_{IR}$ is coherent.

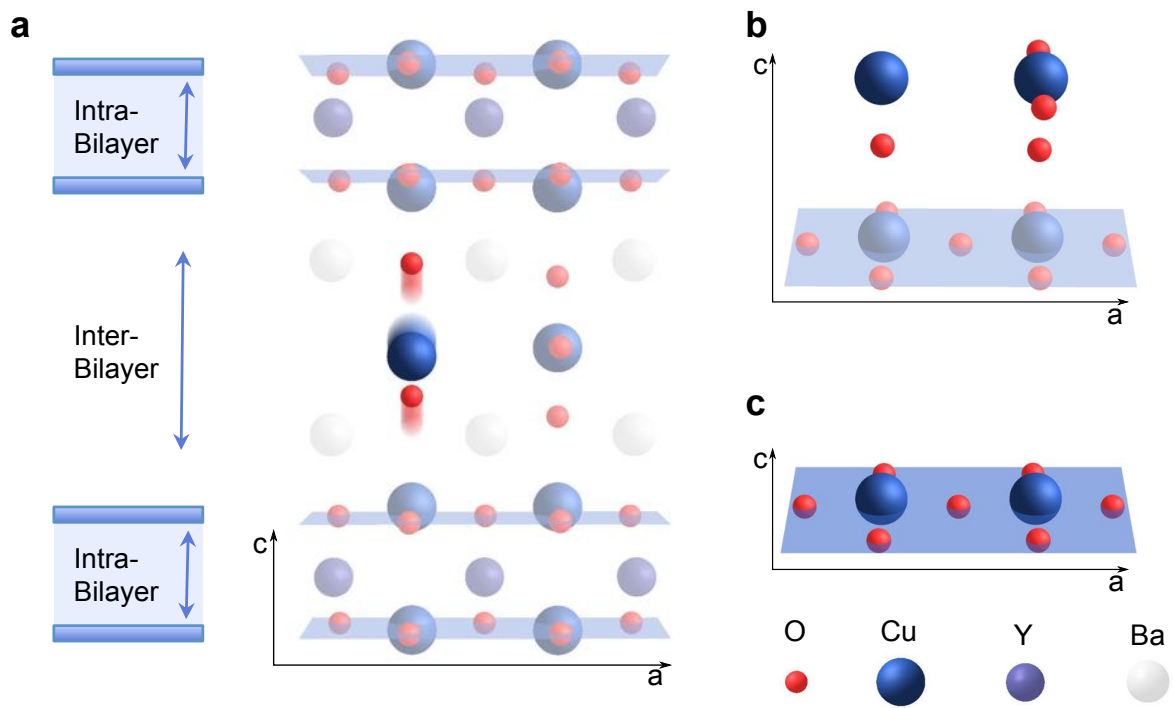

**Figure 2 | Structure of YBa$_2$Cu$_3$O$_{6.5}$ (a)** Structure of orthorhombic YBa$_2$Cu$_3$O$_{6.5}$ and motions of the optically excited B$_{1u}$ mode. The schematic on the left shows the two tunneling regions within and between the bilayers. **(b)** Cu-O Chains, which are either filled (right Cu) or empty (left Cu) in the Ortho-II structure. **(c)** Superconducting CuO$_2$ planes (blue).

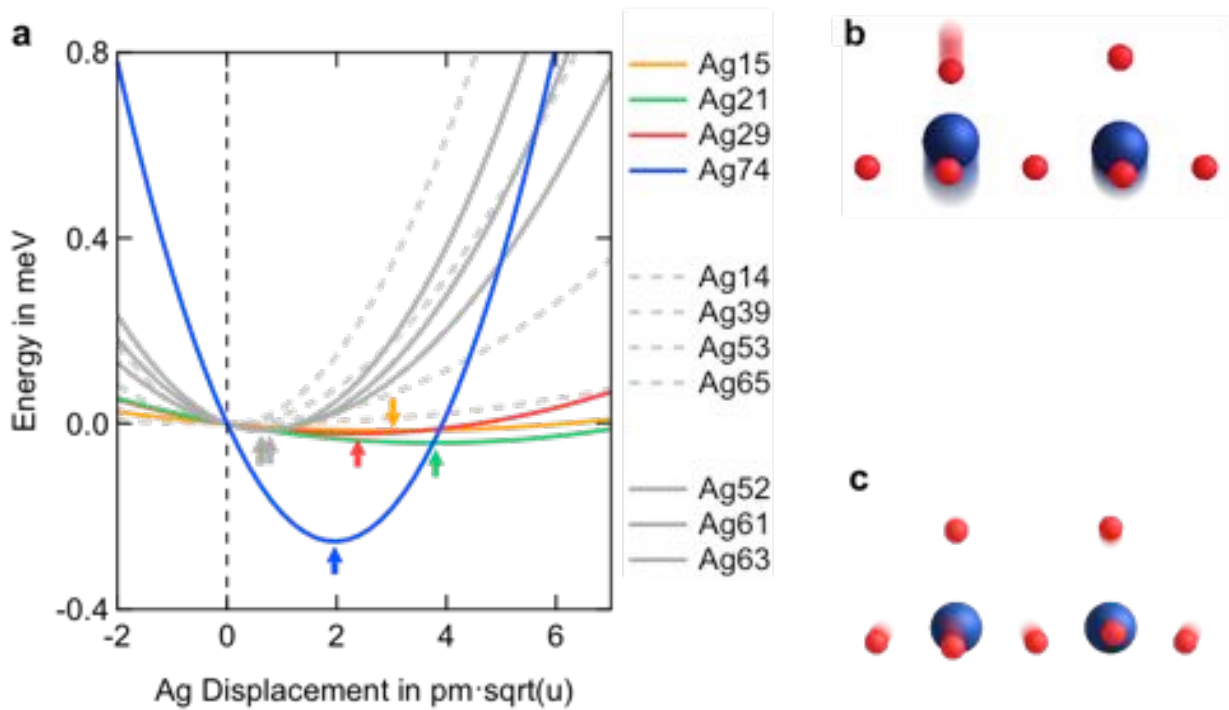

**Figure 3 | First principle calculations of cubic coupling between 11 $A_g$ modes and driven $B_{1u}$ mode. (a)** Energy potentials of all $A_g$ modes for a frozen $B_{1u}$ displacement of $0.14 \text{Å}\sqrt{u}$, corresponding to a change in apical oxygen–copper distance of 2.2pm. The *x*-axis is the amplitude of the $A_g$ Eigenvector (u is the atomic mass unit). Arrows indicate the potential minima. **(b)** Strong coupling occurs to the $A_g(15, 21, 29, 74)$ modes, that comprise a decrease in apical oxygen-copper distance and an increase in in-plane buckling. **(c)** The $A_g(52, 61, 63)$ modes are weakly coupled and govern a breathing motion of the oxygen atoms in the $CuO_2$ plane.

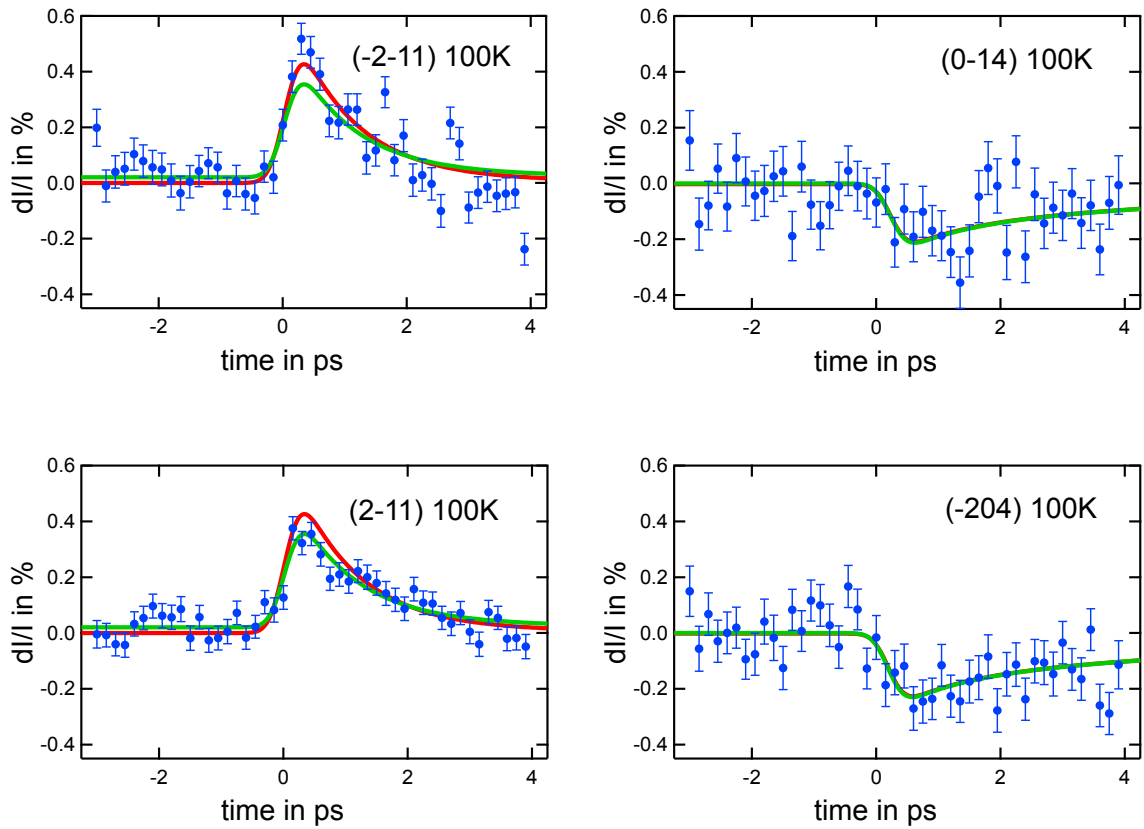

**Figure 4 | Time dependent diffracted peak intensity for four Bragg reflections.** A displacive lattice distortion is observed. The experimental data is fitted (solid curves) by adjusting the $B_{1u}$ amplitude and the relative strength of the two relaxation channels ($\tau_1$=1 ps and $\tau_2$=7 ps) extracted from the optical experiments of Refs. 10 and 11. The relative amplitudes and signs of the curves are determined from the calculated structure using only the four strongest coupled modes (green) or all $A_g$ modes (red). The error bars are extracted from the measurements as 1-σ (67%) confidence interval.

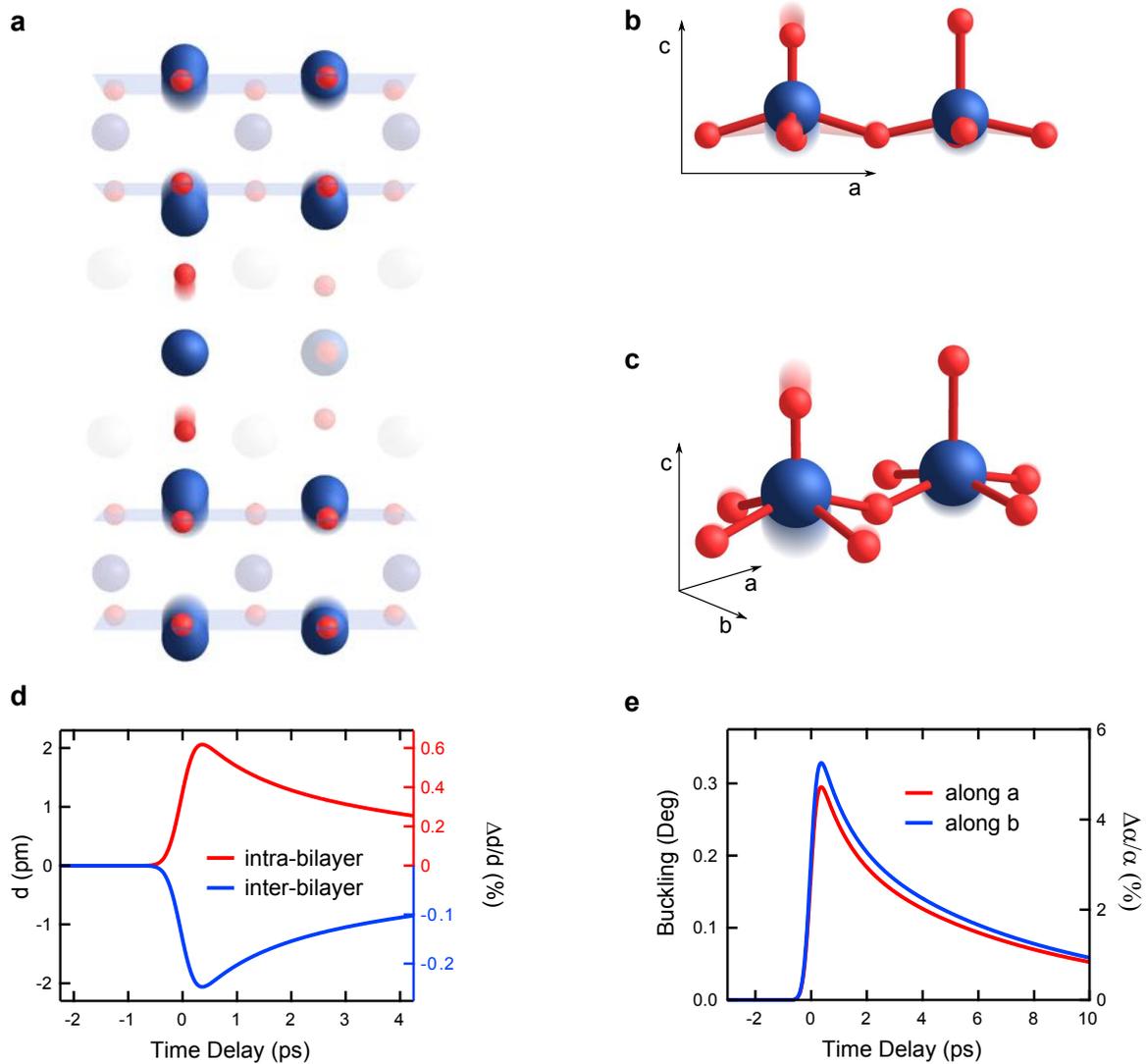

**Figure 5 | Transient lattice structure. (a)** We find a concerted displacive lattice distortion **(b, c)** with a decrease in the apical oxygen – Cu distances by 2.4pm at oxygen deficient sites and an increase in O-Cu-O buckling. **(d)** The intra-bilayer distance increases, while the inter-bilayer distance decreases. Here, the copper atoms of the planes at oxygen deficient chain sites (left hand side in **(a)**) are used to define the position of the planes. **(e)** The in plane O-Cu-O buckling increases by 5% both along *a* and *b* at oxygen deficient sites.

# Online-Only Methods

**Experimental Details**

The x-ray diffraction measurements were carried out with 6.7 keV pulses at the X-ray Pump Probe (XPP) beamline of the Linac Coherent Light Source (LCLS). The energy of the x-rays was selected using a channel-cut Si (111) monochromator with a resolution of 1 eV. The diffraction from each pulse was recorded individually without averaging using a diode. Shot-to-shot normalization to the intensity monitor after the monochromator was used to correct the detected signals for intensity and wavelength fluctuations of the x-ray pulses. The experiment was carried out in grazing incidence geometry with an angle of 5° between the x-rays and the sample surface.

The YBa$_2$Cu$_3$O$_{6.5}$ sample was excited with mid-infrared pulses pulses of ~300 fs duration, generated by optical parametric down-conversion and difference-frequency generation of near-infrared pulses from a Titanium-Sapphire laser. These pulses were tuned to 15μm wavelength with 2μm bandwidth, chosen to be in resonance with the B$_{1u}$ phonon mode. The measurement was carried out at a repetition rate of 120 Hz, while the repetition rate of the mid-infrared pulses was set to 60 Hz. This allowed measuring the equilibrium and excited state for each time delay to correct for any drifts of the Free Electron Laser.

**Structure Factor Calculations**

To deduce the amplitudes $Q_i$ of the atomic displacements along the eigenvectors $\epsilon_i$ of the A$_g$ coordinates from the changes in scattered intensity $I \sim |F|^2$ we calculated the corresponding modulation of the structure factors $F$. Specifically, we quantified $F = \sum_j f_j \exp(-i\boldsymbol{G} \cdot \boldsymbol{r}_j)$, with $\boldsymbol{G}$ being the reciprocal lattice vector of the corresponding diffraction peak, $f_j$ the atomic scattering factor and $\boldsymbol{r}_j$ the positions of the $j$-th atom in the unit cell. By calculating the structure factors for the equilibrium atomic positions $\boldsymbol{r}_j = \boldsymbol{r}_j^{(0)}$ and the transient structure $\boldsymbol{r}_j' = \boldsymbol{r}_j^{(0)} + \sum_i Q_i \epsilon_{ji}$, the relative change $\Delta I$ in diffracted intensity was evaluated as $\Delta I / I = \left( |F(\boldsymbol{r}_j')|^2 - |F(\boldsymbol{r}_j)|^2 \right) / |F(\boldsymbol{r}_j)|^2$.

The changes in signal amplitude are calculated for A$_g$ amplitudes $Q_i$ as predicted by density functional theory calculations for a certain infrared amplitude $Q_{B1u}$ and compared with the experimental findings to determine the quantitative values.

**Density functional theory calculations**

The phonon modes and the nonlinear phonon couplings were obtained using density functional theory calculations with plane-wave basis sets and projector augmented wave pseudopotentials[20,21] as implemented in the VASP software package[22]. The local density approximation was used for the exchange and correlations. We used a cut-off energy of 950 eV for plane-wave expansion and a $4 \cdot 8 \cdot 4$ $k$-point grid for the Brillouin zone integration during self-consistency. We used the experimental lattice parameters of the YBa$_2$Cu$_3$O$_{6.5}$ Ortho-II structure, but relaxed the internal coordinates. The interatomic force constants were calculated using the frozen-phonon method[23], and the PHONOPY software package was used to calculate the phonon frequencies and normal modes[24]. After the normal modes were identified, total energy calculations were performed as a function of infrared $Q_{IR}$ and Raman phonon mode $Q_R$ amplitudes to obtain the energy surfaces. The non-linear coupling between the IR and Raman modes were obtained by fitting the energy surfaces to the polynomial:

$$H = \frac{1}{2}\omega_{IR}^2 Q_{IR}^2 + \frac{1}{2}\omega_R^2 Q_R^2 - a_{21}Q_{IR}^2 Q_R$$

**Coupling strengths of the A$_g$ to the B$_{1u}$ modes**

The energy potential of an A$_g$ Raman mode in the presence of a cubic non-linear coupling to an infrared mode is $V_R = \frac{1}{2}\omega_R^2 Q_R^2 - a_{21}Q_{IR}^2 Q_R$. At equilibrium, when the IR mode is not excited, the potential of the Raman mode has a minimum at $Q_R$ = 0 because the structure is stable at equilibrium. However, when the IR mode is excited externally, the minimum of the Raman mode shifts by an amount $a_{21}Q_{IR}^2/\omega_R^2$. The minima of the A$_g$ modes energy potentials as obtained from DFT calculations for a frozen displacement of the B$_{1u}$ mode of 0.14Å$\sqrt{u}$ are reported in Extended Data Table 1. The mode displacements are given in amplitudes $Q_\alpha$ of the dimensionless eigenvectors of the mode and have the unit Å$\sqrt{u}$, with $u$ being the atomic mass unit. The atomic displacements due to an amplitude $Q_\alpha$ of a mode is given by $\vec{U}_j = \frac{Q_\alpha}{\sqrt{m_j}} \cdot \vec{\omega}_j^\alpha$, where $\vec{U}_j$ is the displacement of the $j$-th atom, $m_j$ is the mass of this atom, and $\vec{\omega}_j^\alpha$ is the corresponding component of the normal-mode vector. Note that $\vec{\omega}_j^\alpha$ is normalized and dimensionless.

**Transient crystal structure**

The static crystal structure of YBa$_2$Cu$_3$O$_{6.5}$ (Ortho-II) is given in Extended Data Table 2. The lattice constants are (a = 7.6586 Å, b = 3.8722 Å, c = 11.725 Å) as determined by single crystal x-ray diffraction at 100 K. The light induced displacements of the atomic positions at peak change in diffracted intensity are reported in Extended Data Table 2.

**Changes to the electronic structure**

The changes of the electronic structure due to the light-induced distortions were studied using the generalized full-potential method within the local density approximation as implemented in the WIEN2k package[25]. The muffin-tin radii of 2.35, 2.5, 1.78, and 1.53 bohrs were used for Y, Ba, Cu, and O, respectively, and a $20 \cdot 40 \cdot 12$ $k$-point grid was used for the Brillouin zone integration. The plane wave cutoff was set at $RK_{max} = 7.0$, where $K_{max}$ is the plane wave cutoff and $R$ is the small muffin-tin radius, i.e. 1.53 bohrs. The density of states (DOS) was generated with a $32 \cdot 64 \cdot 32$ $k$-point grid. Calculations are presented for the equilibrium structure and the transient displaced structure for three B$_{1u}$ amplitudes: the amplitude 0.3 Å$\sqrt{u}$ as determined here, the amplitude 0.8 Å$\sqrt{u}$ as estimated in the experiment of reference 10 and 11 as well as a larger amplitude of 1.2 Å$\sqrt{u}$.

Our calculated electronic structure of the equilibrium YBa$_2$Cu$_3$O$_{6.5}$ (ortho-II), shown in Figure 2 of the main text, is similar to the one calculated previously[26, 27]. The bands near the Fermi level are derived from the Cu $3d$ states from both the planes and chains. The four bands that disperse highly along the path $X - S - Y - \Gamma$ are due to the planar Cu $d_{x^2-y^2}$ states. The broad band that disperses very little along $S - Y$ is due to the Cu $d_{z^2}$ states from the filled chain. The oxygen deficient chains that control the hole doping give rise to fairly flat electronic bands with dominant Cu $d_{xz}$ and $d_{yz}$ character that are very close to the Fermi level at the $Y$ point. The electronic structure calculations predict some hybridization between these bands and the planar Cu bands that creates an anti-crossing near $Y$ In the equilibrium structure, this anti-crossing is close to the Fermi level, giving rise to pockets with unfilled chain Cu character in the Fermi surface. (see Extended Data Figures 4 and 5).

The displacements due to the nonlinear couplings cause noticeable changes to the electronic structure around the Fermi level. There are three main effects:

(i) The light-induced displacements reduce the width of the planar Cu bands, which leads to an increase in the planar Cu contribution to the DOS at the Fermi level.
(ii) The atomic displacements cause a charge transfer from the planes to the oxygen deficient chains. As the unfilled Cu chain bands lower in energy and move below the Fermi level with increasing light-induced displacements, the planar Cu states increase in energy, becoming more unoccupied. I.e., there is an effective hole doping of the planar Cu states due to the light-induced displacements (see Extended Data Figure 6).

(iii) The changes in the relative occupations of the bands also cause a topological change in the Fermi surface. The light-induced displacements increase the filling of the unfilled chain Cu bands, which decreases the size of the pockets in the Fermi surface. Above a threshold of $0.8\ \text{Å}\sqrt{u}$, the oxygen deficient chain Cu bands become fully filled and the Fermi surface consists solely of two-dimensional planar Cu and one-dimensional filled chain Cu sheets.

**Quartic order coupling**

To verify that the nonlinear phonon coupling is dominated by third order, as discussed in the main text, we checked for signals from the next (fourth) order, described by the term $Q_{IR}^2 Q_j^2$ in the nonlinear Hamiltonian:

$$H_{NL} = \frac{1}{2}\omega_R^2 Q_j^2 - a_{21} Q_{IR}^2 Q_j - a_{22} Q_{IR}^2 Q_j^2$$

As noted in the text, when the directly driven infrared mode is of $B_{1u}$ symmetry, third order coupling is nonzero only to modes of $A_g$ symmetry. However, coupling to any mode $Q_j$ is allowed through $Q^2 Q^2$ coupling, in particular to in-plane $B_g$ modes.

Note first that small amplitude $B_{1u}$ excitations would simply renormalize the frequency of a second mode $Q_j$. This can directly be deduced from the equation of motion, where the driving force is given by the coupling term $a_{22} Q_{IR}^2 Q_j$, which is linear in $Q_j$.

$$\ddot{Q}_j + 2\gamma_j Q_j \dot{Q}_j + \omega_j^2 Q_j = 2 a_{22} Q_{IR}^2 Q_j$$

Upon $Q_{IR}$ displacement, the anharmonically coupled mode experiences a renormalization of its frequency $\omega_j' = \omega_j \sqrt{1 - 2 a_{22} Q_{IR}^2 / \mu_j}$.

However, above a threshold-amplitude $Q_{IR}^*$, the frequency of the second mode $Q_j$ becomes imaginary[3] and the lattice becomes unstable. Importantly, such instability can take place in two directions, depending on the *random* instantaneous state of the system (mode amplitude $Q_j$ and its velocity $dQ_j/dt$. This manifests in a change from a parabolic to a double well energy potential as shown in the Extended Data Figure 1.

Hence, fourth order effects need to be identified by analyzing the diffraction of each individual x-ray pulse, whereas the unsorted average is expected to be zero even if quartic coupling is sizeable. In the experiment, we sorted all positive and negative deviations from the average signal of all shots to obtain the $Q^2Q^2$ response at a specific time delay. Averaging them separately and subtracting negative from positive deviations then gives the intensity changes from $Q^2Q^2$ only.

Time-resolved x-ray diffraction was measured for four Bragg reflections, sensitive to $A_g$ and to $B_{2g}$ displacements. The results of these experiments are shown in Extended Data Figure 2.

Within our resolution, we find no evidence of quartic contributions. The amplitude of the infrared motion is below the threshold beyond which fourth order coupling induces lattice displacements.

# References Online-Only Methods

# Extended Data Figure and Table Legends

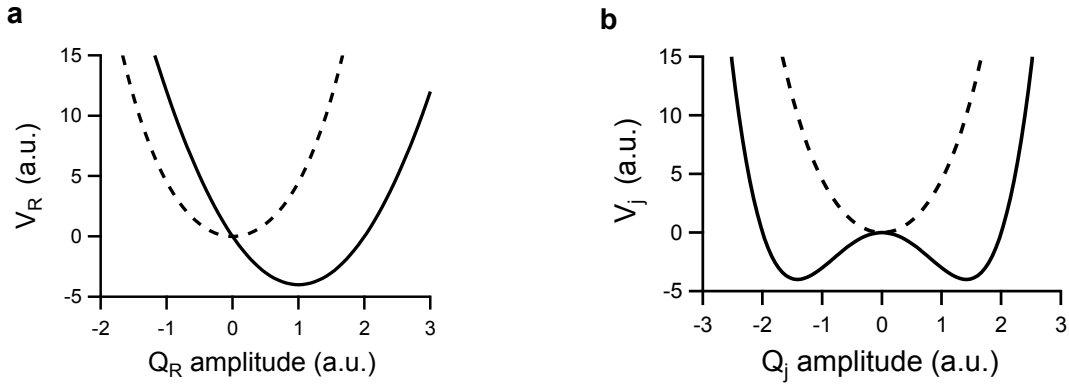

**Extended Data Figure 1 | Nonlinear lattice dynamics in the limit of cubic and quartic coupling.** Dashed lines: Potential energy of a mode $Q$ as a function of mode amplitude. **a)** A static distortion $Q_{IR}^*$ shifts the potential of all modes $Q_R$ that are coupled through a $Q_{IR}^2 Q_R$ coupling, (solid line) displacing the equilibrium position towards a new minimum. **b)** Due to quartic $Q_{IR}^2 Q_j^2$ coupling, the energy potential of a coupled mode $Q_j$ is deformed symmetrically upon static distortion $Q_{IR}^*$. The frequency of the mode first softens until it is destabilized, which manifests in a double well potential (solid line).

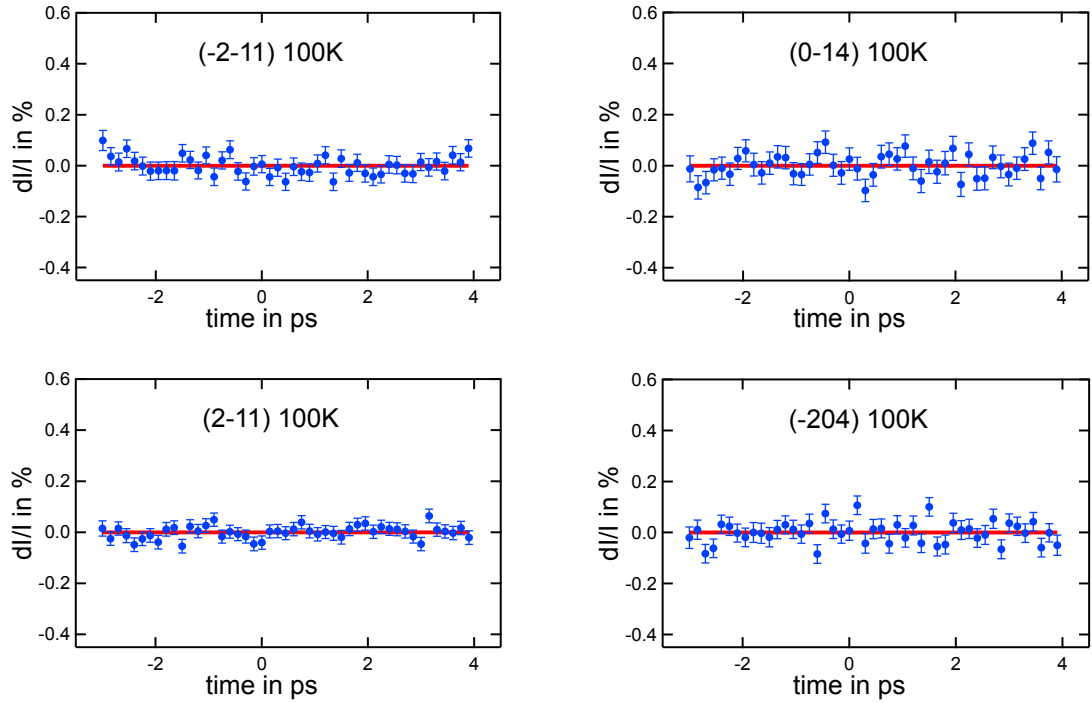

**Extended Data Figure 2 | Changes in diffracted intensity of specific Bragg reflections from fourth order coupling for different time delays between pump and probe pulse.** We find no evidence of lattice distortions originating from fourth order contributions to the phonon coupling. The amplitude of the infrared mode $Q_{IR}$ is below the threshold beyond which fourth order effects destabilize coupled phonon modes. The error bars are extracted from the measurements as 1-σ (67%) confidence interval.

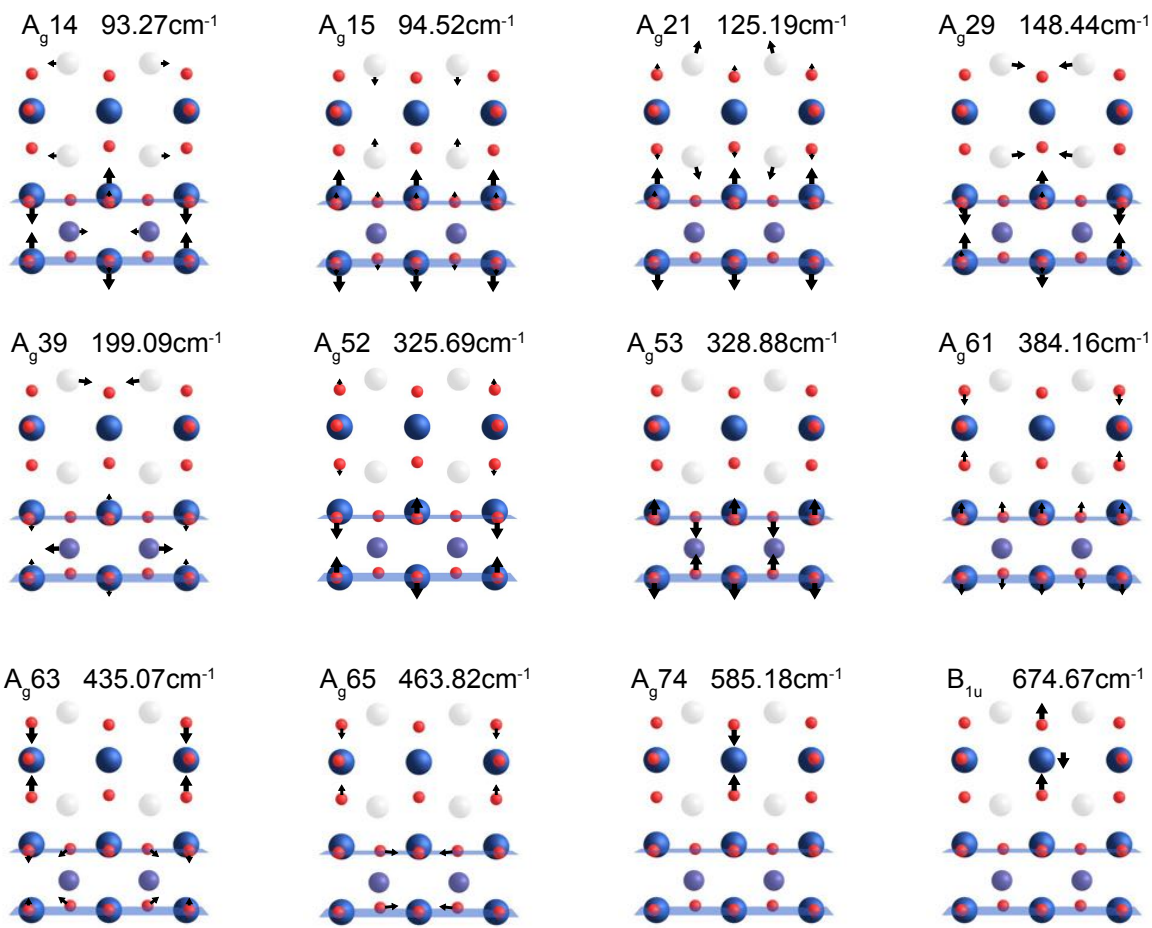

**Extended Data Figure 3 | Phonon modes of Ortho-II YBa$_2$Cu$_3$O$_6.5$.** Shown are sketches of the resonantly excited B$_{1u}$ mode and all 11 A$_g$ modes to which the coupling strengths (reported in Extended Data Table1) have been calculated.

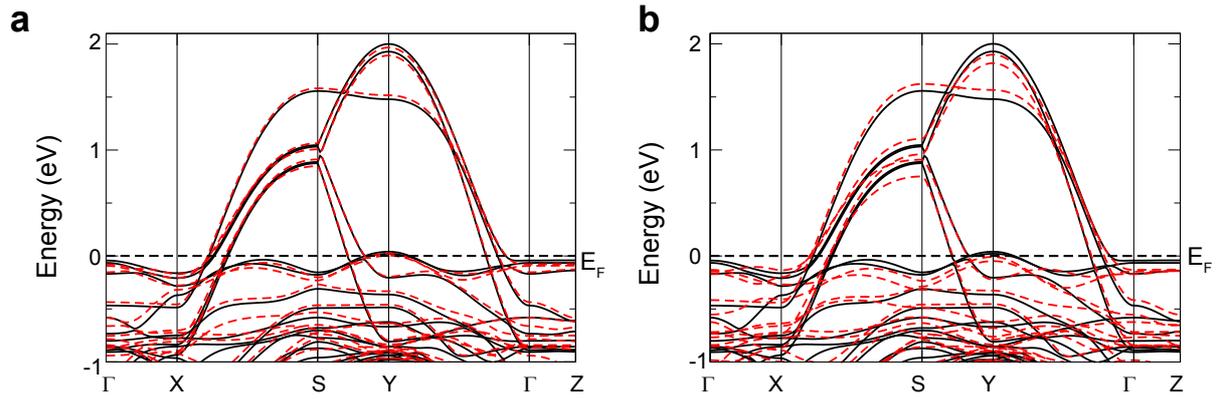

**Extended Data Figure 4 | Band structure of the equilibrium (black line) and transient crystal structure.** The band structure is plotted along $\Gamma(0,0,0) \rightarrow X(0.5,0,0) \rightarrow S(0.5,0.5,0) \rightarrow Y(0,0.5,0) \rightarrow \Gamma(0,0,0) \rightarrow Z(0,0,0.5)$ **a)** for $0.8 \text{ Å}\sqrt{u}$, which is the amplitude estimated for the geometry of reference 10 and 11 and **b)** $1.2 \text{ Å}\sqrt{u}$ $B_{1u}$.

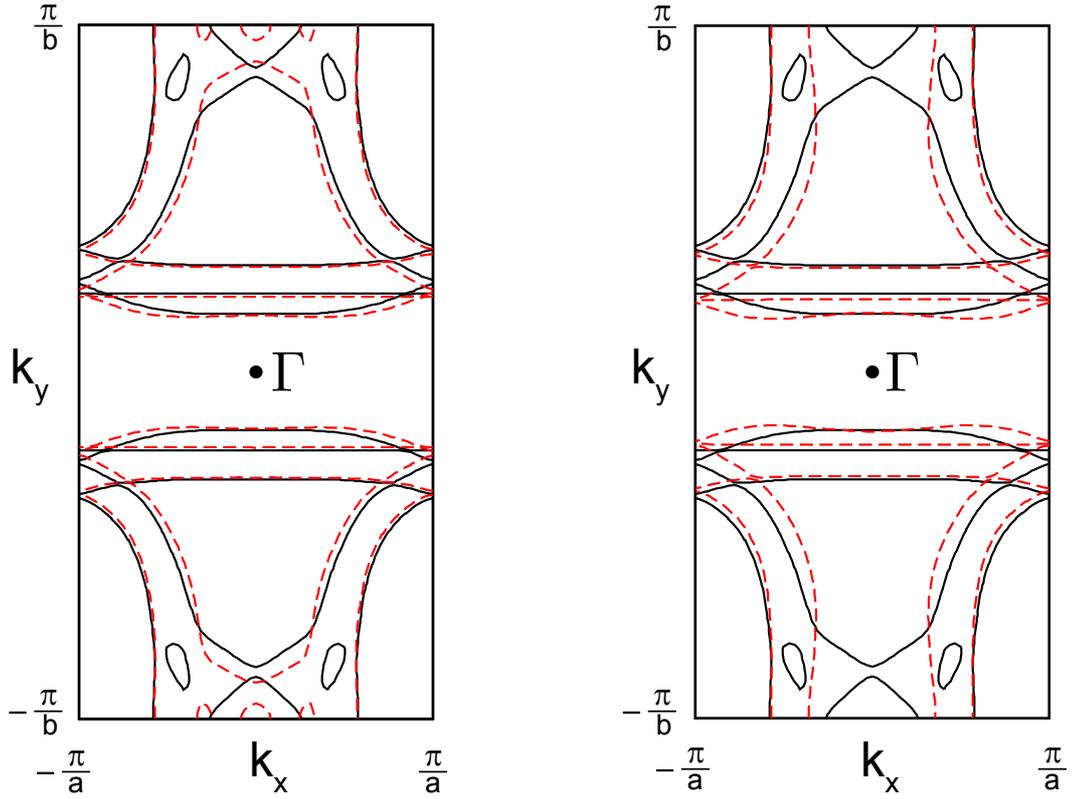

**Extended Data Figure 5 | $k_z=0$ cut of the Fermi surface of the equilibrium (black line) and transient crystal structure (red dashed line).** In the equilibrium structure, the bands of the unfilled chain Cu give rise to pockets in the Fermi surface. The light induced displacements shift the density of states of these bands to lower energies, increasing the filling and reducing the pockets. Above a threshold of 0.8 Å$\sqrt{u}$, the oxygen deficient chain bands become fully filled, the pockets close and the Fermi surface consists solely of two-dimensional planar Cu and one-dimensional filled chain states. The Fermi surface is shown in the displaced state for **b)** 0.8 Å$\sqrt{u}$ and **b)** 1.2 Å$\sqrt{u}$ $B_{1u}$ amplitude.

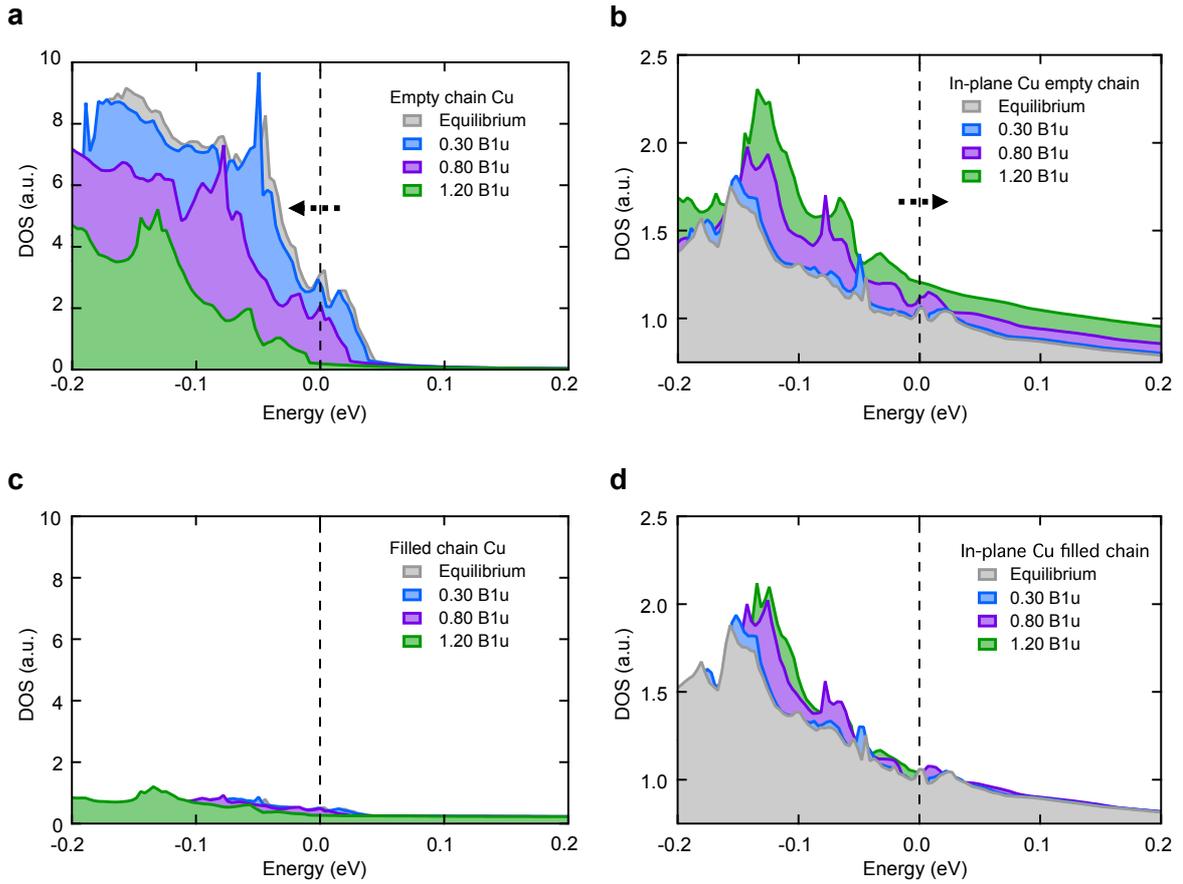

**Extended Data Figure 6 | Changes in the density of states in the CuO₂ plane and the Cu–O chains.** These are obtained from a projection of the density of states onto the copper muffin-tin spheres. **a**, **b**, In the light-induced state, the density of states of the O-deficient chain lowers in energy (**a**), whereas the opposite effect is observed for the Cu in the plane below (**b**). This corresponds to charge transfer from the planes to the chains. **c**, **d**, The density of states of the filled chain Cu is not strongly affected (**c**). The bands of the planar Cu atoms narrow, which leads to an increase in the density of states near the Fermi level both at sites with filled (**d**) and empty chains (**b**). The effect is already visible for a $Q_{B1u}$ amplitude of $0.3\,\text{Å}\sqrt{u}$ (blue) but becomes more prominent for larger displacements of $0.8\,\text{Å}\sqrt{u}$ (purple) and $1.2\,\text{Å}\sqrt{u}$ (green).

| Mode | Displacement (Å√u) |
|---|---|
| Ag14 | -0.002 |
| Ag15 | 0.031 |
| Ag21 | -0.038 |
| Ag29 | -0.023 |
| Ag39 | 0.000 |
| Ag52 | 0.007 |
| Ag53 | 0.000 |
| Ag61 | -0.007 |
| Ag63 | 0.007 |
| Ag65 | -0.001 |
| Ag74 | 0.020 |

**Extended Data Table 1 | Mode displacements** $(Å\sqrt{u})$ Energy potential minima of the $A_g$ modes as obtained from DFT calculations for a frozen displacement of the $B_{1u}$ mode of $0.14\ Å\sqrt{u}$, which corresponds to a change in apical oxygen–copper distance of 2.2pm.

| Atom | Equilibrium Structure (Å) | | | Displacements (pm) 0.3Å√u | | | Displacements (pm) 0.8Å√u | | | Displacements (pm) 1.2Å√u | | |
|---|---|---|---|---|---|---|---|---|---|---|---|---|
| | x | y | z | x | y | z | x | y | z | x | y | z |
| Y | 1.922 | 1.936 | 5.863 | 0.111 | 0.000 | 0.000 | 0.769 | 0.000 | 0.000 | 1.835 | 0.000 | 0.000 |
| Y | 5.737 | 1.936 | 5.863 | -0.111 | 0.000 | 0.000 | -0.769 | 0.000 | 0.000 | -1.835 | 0.000 | 0.000 |
| Ba | 1.861 | 1.936 | 2.224 | 0.159 | 0.000 | 0.027 | 0.142 | 0.000 | 0.040 | 0.167 | 0.000 | 0.227 |
| Ba | 5.797 | 1.936 | 9.501 | -0.159 | 0.000 | -0.027 | -0.142 | 0.000 | -0.040 | -0.167 | 0.000 | -0.227 |
| Ba | 1.861 | 1.936 | 9.501 | 0.159 | 0.000 | -0.027 | 0.142 | 0.000 | -0.040 | 0.167 | 0.000 | -0.227 |
| Ba | 5.797 | 1.936 | 2.224 | -0.159 | 0.000 | 0.027 | -0.142 | 0.000 | 0.040 | -0.167 | 0.000 | 0.227 |
| Cu | 0.000 | 0.000 | 0.000 | 0.000 | 0.000 | 0.000 | 0.000 | 0.000 | 0.000 | 0.000 | 0.000 | 0.000 |
| Cu | 3.829 | 0.000 | 0.000 | 0.000 | 0.000 | 0.000 | 0.000 | 0.000 | 0.000 | 0.000 | 0.000 | 0.000 |
| Cu | 0.000 | 0.000 | 4.227 | 0.000 | 0.000 | -0.376 | 0.000 | 0.000 | -0.799 | 0.000 | 0.000 | -1.659 |
| Cu | 0.000 | 0.000 | 7.498 | 0.000 | 0.000 | 0.376 | 0.000 | 0.000 | 0.799 | 0.000 | 0.000 | 1.659 |
| Cu | 3.829 | 0.000 | 4.231 | 0.000 | 0.000 | -1.032 | 0.000 | 0.000 | -4.999 | 0.000 | 0.000 | -11.451 |
| Cu | 3.829 | 0.000 | 7.494 | 0.000 | 0.000 | 1.032 | 0.000 | 0.000 | 4.999 | 0.000 | 0.000 | 11.451 |
| O | 0.000 | 1.936 | 0.000 | 0.000 | 0.000 | 0.000 | 0.000 | 0.000 | 0.000 | 0.000 | 0.000 | 0.000 |
| O | 1.915 | 0.000 | 4.443 | -0.055 | 0.000 | 0.066 | -0.233 | 0.000 | 0.293 | -1.024 | 0.000 | 1.144 |
| O | 5.743 | 0.000 | 7.282 | 0.055 | 0.000 | -0.066 | 0.233 | 0.000 | -0.293 | 1.024 | 0.000 | -1.144 |
| O | 1.915 | 0.000 | 7.282 | -0.055 | 0.000 | -0.066 | -0.233 | 0.000 | -0.293 | -1.024 | 0.000 | -1.144 |
| O | 5.743 | 0.000 | 4.443 | 0.055 | 0.000 | 0.066 | 0.233 | 0.000 | 0.293 | 1.024 | 0.000 | 1.144 |
| O | 0.000 | 1.936 | 4.431 | 0.000 | 0.000 | -0.015 | 0.000 | 0.000 | -0.076 | 0.000 | 0.000 | 0.215 |
| O | 0.000 | 1.936 | 7.294 | 0.000 | 0.000 | 0.015 | 0.000 | 0.000 | 0.076 | 0.000 | 0.000 | -0.215 |
| O | 3.829 | 1.936 | 4.440 | 0.000 | 0.000 | 0.127 | 0.000 | 0.000 | 0.490 | 0.000 | 0.000 | 1.690 |
| O | 3.829 | 1.936 | 7.285 | 0.000 | 0.000 | -0.127 | 0.000 | 0.000 | -0.490 | 0.000 | 0.000 | -1.690 |
| O | 0.000 | 0.000 | 1.857 | 0.000 | 0.000 | -0.057 | 0.000 | 0.000 | 0.199 | 0.000 | 0.000 | -0.382 |
| O | 0.000 | 0.000 | 9.868 | 0.000 | 0.000 | 0.057 | 0.000 | 0.000 | -0.199 | 0.000 | 0.000 | 0.382 |
| O | 3.829 | 0.000 | 1.758 | 0.000 | 0.000 | 1.335 | 0.000 | 0.000 | 7.355 | 0.000 | 0.000 | 12.342 |
| O | 3.829 | 0.000 | 9.967 | 0.000 | 0.000 | -1.335 | 0.000 | 0.000 | -7.355 | 0.000 | 0.000 | -12.342 |

**Extended Data Table 2 | Equilibrium structure of YBa$_2$Cu$_3$O$_{6.5}$ and light induced displacements**. Atomic positions are given in Ångström (x,y,z), the light-induced atomic displacements in pm.